# INVESTIGATIONS OF THE INFLUENCES OF A CNN'S RECEPTIVE FIELD ON SEGMENTATION OF SUBNUCLEI OF BILATERAL AMYGDALAE

Han Bao[1]

[1]School of Computer Science, University of Nottingham Ningbo China, Ningbo, China

***ABSTRACT***

*Segmentation of objects with various sizes is relatively less explored in medical imaging, and has been very challenging in computer vision tasks in general. We hypothesize that the receptive field of a deep model corresponds closely to the size of object to be segmented, which could critically influence the segmentation accuracy of objects with varied sizes. In this study, we employed "AmygNet", a dual-branch fully convolutional neural network (FCNN) with two different sizes of receptive fields, to investigate the effects of receptive field on segmenting four major subnuclei of bilateral amygdalae. The experiment was conducted on 14 subjects, which are all 3-dimensional MRI human brain images. Since the scale of different subnuclear groups are different, by investigating the accuracy of each subnuclear group while using receptive fields of various sizes, we may find which kind of receptive field is suitable for object of which scale respectively. In the given condition, AmygNet with multiple receptive fields presents great potential in segmenting objects of different sizes.*

## 1. INTRODUCTION

Last few years have witnessed convolutional neural networks (CNNs) [4, 5, 6] gaining great success in computer vision area [2], e.g. [4,7]. While CNN has appeared for a long period of time [8], a typical application of CNN is in classification [7, 9, 15, 16, 17, 18]. Except classification, object detection is also an applicable area of CNN [4, 5, 6, 19, 20, 21]. Apart from classification and detection, another critical area of applying CNN is segmentation, in which Markov Random Field (MRF) or Conditional Random Field (CRF) has gained significant successes [22]. Ronneberger et al. [9] came up with an encoder-decoder architecture called U-Net, which is used to address the problem in segmentation. The encoder part compresses the input images into lower-resolution feature maps via downsampling or pooling layers, and the decoder part aims to recover the full-resolution label map from these feature maps for pixel-to-pixel semantic classification. However, such design could pose a few problems when segmenting structures with small spatial extent [2]. Due to much feature loss during a large amount of downsampling, U-Net may overlook some small objects, which may lead to inaccurate segmentation [23]. While in medical research area, a lot of objects need to be segmented are small, one of which is amygdalae [24]. In this case, U-Net and many other networks are no longer suitable. To handle this issue, several experiments have been conducted. At first, image pyramid method [10] and feature pyramid method [11] are put forward. The former one using multiple scales of images as input, and in each scale, object segmentation can be performed respectively. Segmentation accuracy can be improved by this way. However, object segmentation is performed on various scales of images, which means inference time may be increased. Due to this issue, image pyramid method seems unusable and unpractical [2]. The later one, feature pyramid method, avoids the drawback of image pyramid method by fusing multi-scale feature maps in different layers of CNN. [11] first manifested this concept. By inserting feature channels from adjacent levels, fast feature pyramid can be constructed. Inferring from these feature maps, rather than multi-scaled images, can reduce computation cost. One application of this method is FPN [12], which uses an augmented top-down pathway and horizontal connections to merge strong semantic information in high-level features [2]. Nevertheless, another issue occurs. Since features are elicited from different layers, the inference results of objects in different scales are still of difference [2]. Due to this problem, feature pyramid is not an ideal alternative method of image pyramid.

In above to methods, image pyramid utilizes more real images for different scales, although it is not efficient. Feature pyramid method discards some extent of consistency of images in different scales to save inference time. However, the result is not in a high level. To make further improvement, we put forward a dual-branch fully convolutional neural network (FCNN) – AmygNet. By implementing different sizes of receptive fields in one network, we make segmenting small-scale objects more efficiently.

Since the involvement of computer vision, automatic monitoring may be realistic in biomedical researches [13], and much attention has been drawn in medical imaging in recent years [14]. It seems implementing object segmentation with high accuracy may have a meaningful and promising future. However, samples used in medical research are different from normal 2-dimensional images, they are 3-dimensional and generated by medical instruments. Under this condition, the segmentation accuracy may be varied using normal network. In addition, unlike a wealth of labeled images taken in our normal life time, labeled samples generated in medical study are relatively rare, which means it is not easy to find an appropriate dataset to utilize in the experiment, so getting a higher segmentation accuracy rate using a limited number of samples are significant. Also, segmenting small-scale objects like amygdalae is challenging [24]. Segmenting some other relatively larger medical objects has been performed by a great deal of researchers, such as segmenting brain [25, 26, 27, 29], lung [28, 30] and breast [31, 32]. However, until now no study is focused on segmenting subnuclear groups of amygdalae, as far as we know. Thus, research on segmenting subnuclear groups of amygdalae is of great significance and worth a try. Fortunately, we obtain a set of 3-dimensional labeled samples of human brain as our dataset. In our experiment, we introduce a novel network – Amyg Network (AmygNet) to segment amygdalae in a whole human brain. By segmenting different subnuclear parts in different scales, AmygNet is able to infer the integrated amygdala, and during this process, AmygNet may show its great potential in segmenting small objects and handling various scales of objects.

## 2. RELATED WORK

In this section, we will present some work conducted before the formal experiment, including study on receptive field and two pilot experiments.

### 2.1.About Receptive Field

Receptive field is one of the most fundamental and significant components in a convolutional neural network (CNN). Feature map of each layer of CNN is generated via receptive field – every single pixel in feature map is generated by receptive field, so it is not overstated that one trivial change in receptive field may cause consecutive influence on the whole network structure. However, the study of receptive field in previous work is relatively rare [2].

In our experiment, we are concerned about the size of receptive field. To obtain various sizes of receptive field, one option is to increase the length of side of convolutional kernel directly. However, one primary drawback of this transformation is that the need of computational resource increases heavily, which may lead to the dramatic increase on experiment time. To alleviate this problem, we implement an alternate scheme – dilated convolution (aka atrous convolution) [3]. Instead of directly enlarging the size of convolutional kernel, we use a new parameter, dilation rate, to control the enlargement of convolutional kernel, furtherly enlarge the size of receptive field, without increasing the amount of computation. For a dilated convolution with dilation rate $d$, $d-1$ zeros are inserted between consecutive filter values, making kernel size larger without aggrandizing the number of arguments and computation [2]. For instance, a dilated 3×3×3 convolution can represent the convolution with kernel size of $3+2(d-1)$, doing the same amount of calculation as 3×3×3 convolution does.

Actually, only the size of receptive field of the first convolution layer's feature map equals to the size of its filter, or kernel. The size of deeper layers' receptive fields is related to previous layers' filters size and strides. To calculate the size of receptive fields $l_k$ of layer $k$, a formula can be used:

$$l_k = l_{k-1} + ((f_k - 1) * \prod_{i=1}^{k-1} s_i)$$

where $l_{k-1}$ represents the receptive field of layer $k-1$, $f_k$ is the filter size (length, width or height, assuming they are the same), and the stride of layer $i$ is $s_i$. Since the size of a receptive field is related to all the previous layers (accumulative), the speed of increase of receptive field size is relatively fast, which means receptive field in deeper layer is much larger than previous layers.

## 2.2.Pilot Experiments

In order to have a general understanding of the relation between segmentation results and the size of receptive field, we conduct two pilot experiments first, one using a network with normal (non-dilated) receptive field called Normal AmygNet, while another using a network with dilated receptive field called Dilated AmgyNet. Four ResNet [21] building blocks (residual connection is implemented in each block) are included consecutively in both AmygNets. The only difference between Normal AmygNet and Dilated AmygNet is the size of receptive field in each weight layer of each ResNet building block. In Normal AmygNet, each FCNN has a standard convolutional kernel inside, with dilated rate of 1, which means receptive fields are non-dilated. However, receptive fields in Dilated AmygNet are not standard – they are all dilated by different dilation rates based on standard convolutional kernel implemented in Normal AmygNet. For example, in first building block of ResNet, two convolutions in two weight layers have dilation rate of 2 and 4 respectively, so they have convolutions with kernel sizes of 5×5×5 and 9×9×9 respectively and receptive fields with sizes of 5 and 13 respectively, which are both bigger than standard convolutional kernel (5×5×5 and 9×9×9 compared with 3×3×3) and receptive fields (5 and 13 compared with 3 and 5 respectively). By only modifying the size of receptive fields, we keep rest parts of two networks the same while doing pilot experiments.

A set of 14 3-dimentional MRI human brain images are selected as our dataset. All 14T1-weighed subjects in our dataset are collected from a GE ME750 3.0 T MRI scanner, which has the product 8-channel head coil. Each volume has a dimension of 171×236×191. A previous study performed four separating imaging sessions on each subject through an amount of half-transforms, as well as co-registered and averaged [1]. Skull-removing, bias-field correcting and intensity normalization in volume are performed during pre-processing process. Images have an original resolution of 240×240×124mm (0.9375×0.9375×1mm). In an AC-PC alignment with landmark-base, they are adjusted into size of 171×236×191 bricks of 1mm isotropic voxels. Rotation in sagittal plane to the “pathological plane” is involved later, in order to match post-mortem atlases. Left and right amygdalae were manually labeled by an amygdalae anatomy expert (BN), as well as four subnuclear groups on each side – lateral, basal, centromedial and cortico-superficial. Thus, we set class number as 8 – 4 subnuclear groups on left side and 4 subnuclear groups on right side. Subnuclear groups are labeled as shown in Table 1.

Table 1. Label numbers of classes in each sample and their corresponding part. In the actual experiment, there are 11 classes in total – 8 classes above plus 2 “leftover” parts (1 in each side) and 1 background part. Since in medical study, researchers normally do not care about “leftover” parts and background part, we discard these 3 classes and leave the above 8 classes.

| Part | Lateral on left side | Basal on left side | Centromedial on left side | Cortico-superficial on left side |
|---|---|---|---|---|
| Class number | 1 | 2 | 3 | 4 |
| Part | Lateral on right side | Basal on right side | Centromedial on right side | Cortico-superficial on right side |
| Class number | 5 | 6 | 7 | 8 |

To investigate the relation between the size of receptive field and the size of object, 8 subnuclear groups are sorted by their sizes. Since left part and right part of an amygdala are symmetrical (each part has same structure), we do not need to separate subnuclear groups into two parts (left and right) while sorting the order of sizes of subnuclear groups. Sorted by volume-to-surface ratio, from the largest to the smallest, the order of subnuclear groups should be lateral, basal, centromedial and cortico-superficial. Thus, we regard the former two subnuclear groups – lateral and basal – as big objects, and the latter two subnuclear groups – centromedial and cortico-superficial – as small objects.

Both Normal AmygNet and Dilated AmygNet are trained in 11 patches Google Colaboratory, using 1 GPU. We set number of classes to 11 (8 classes in Table 1 plus 2 “leftover” classes and 1 background class), although we only concern about 8 classes (4 subnuclear groups in each side of an amygdala, as shown in Table 1). In total, we train 500 epochs. While implementing ResNet, each convolution implements a kernel size of 3×3×3, except reshape convolutions, which have kernel size of 1×1×1 (used for residual connection). Stride in each convolution is same, too, which has a value of 1. Normal AmygNet uses dilation rate of convolution of 1 (non-dilated receptive field) in each convolution manipulation. Dilated AmygNet uses different dilation rates in each convolution. In conv1 stage, dilation rate is 2 and 4 in first and second convolution respectively. In conv2 stage, dilation rate is 2 and 8 in first and second convolution respectively. In conv3 stage, dilation rate is 2 and 4 in first and second convolution respectively. In conv4 stage, dilation rate is 2 and 1 in first and second convolution respectively. In both Normal AmygNet and Dilated AmygNet, before 4 layers of ResNet blocks, there is a convolution layer, which has kernel size of 3×3×3 and stride of 1, outputs 30 channels. This generates the number of input channels of ResNet blocks, which is 30. After going through four times of manipulation using ResNet building block, Normal AmygNet and Dilated AmygNet will output 50 channels when ResNet building block manipulation is over. Then, after 2 full connected layers, it comes to output layer, which will generate final results.

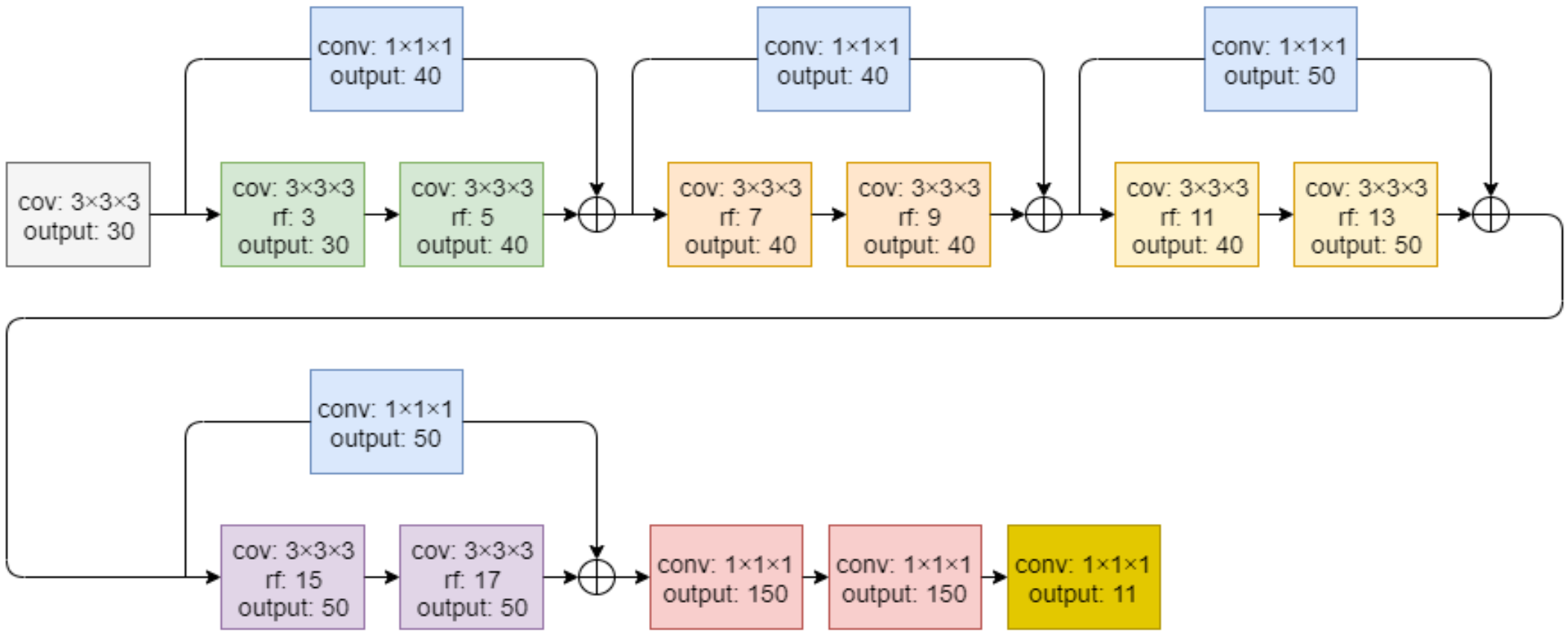

Figure 1. The structure of Normal AmygNet. Four ResNet building blocks are implemented. "conv" represents size of convolutional kernels, while "rf" means the size of receptive field at that layer and "output" shows the output size of that layer. Since convolution size in each layer is 3×3×3, there is no dilation involved in Normal AmygNet. Size of receptive field of last convolution in ResNet building block is 17. First layer and two full connected layers are also shown at the first and last stage of experiment respectively. Except the size of convolutional kernels and receptive fields, other parameter settings are the same as Dilated AmygNet.

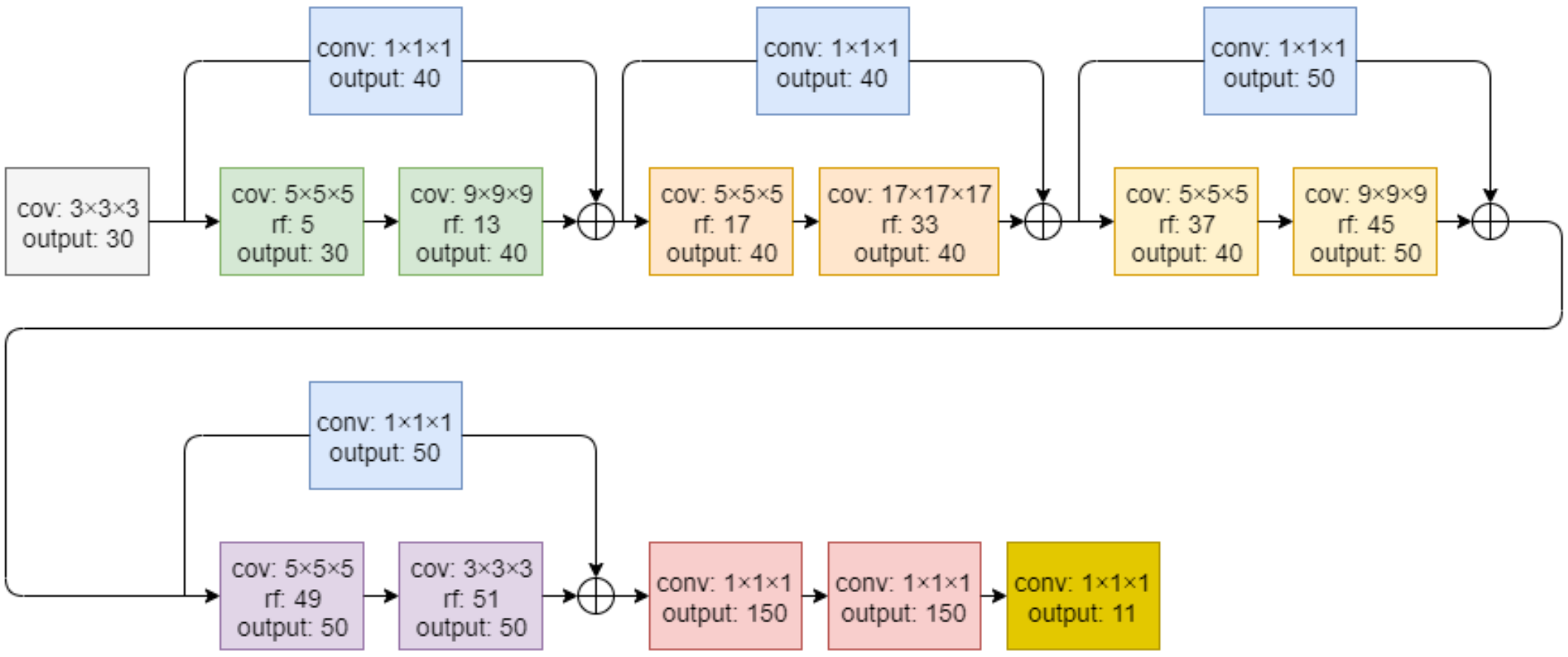

Figure 2. The structure of Dilated AmygNet. Four ResNet building blocks are implemented. "conv" represents size of convolutional kernels, while "rf" means the size of receptive field at that layer and "output" shows the output size of that layer. Since convolution size in each layer is different, different dilation rates are used, which means dilation is involved in each layer (except the last one). Size of receptive field of last convolution in ResNet building block is 51, which is 3 times larger compared with Normal AmygNet. First layer and two full connected layers are also shown at the first and last stage of experiment respectively. Except the size of convolutional kernels and receptive fields, other parameter settings are the same as Dilated AmygNet.

Running detail is the same as AmygNet and will be shown in "Experiment" section. Dice and ASSD are used to evaluate the results. Dice measures the volume overlap of inference and sample, which represents the similarity between testing results and labeled samples (closer to 1, higher similarity). Basically, Dice can measure the accuracy of segmentation result. ASSD measures the boundary overlap of inference and sample, which reflects the localization ability of boundary and regions of interests (closer to 0, better localization ability). Setting each sample as testing sample once, both Normal AmygNet and Dilated AmygNet generate 14 results respectively (since there are 14 samples in total). By averaging the 14 results in each class, and then averaging the results of two classes of the same subnuclear group, we get Figure 3 and Figure 4.

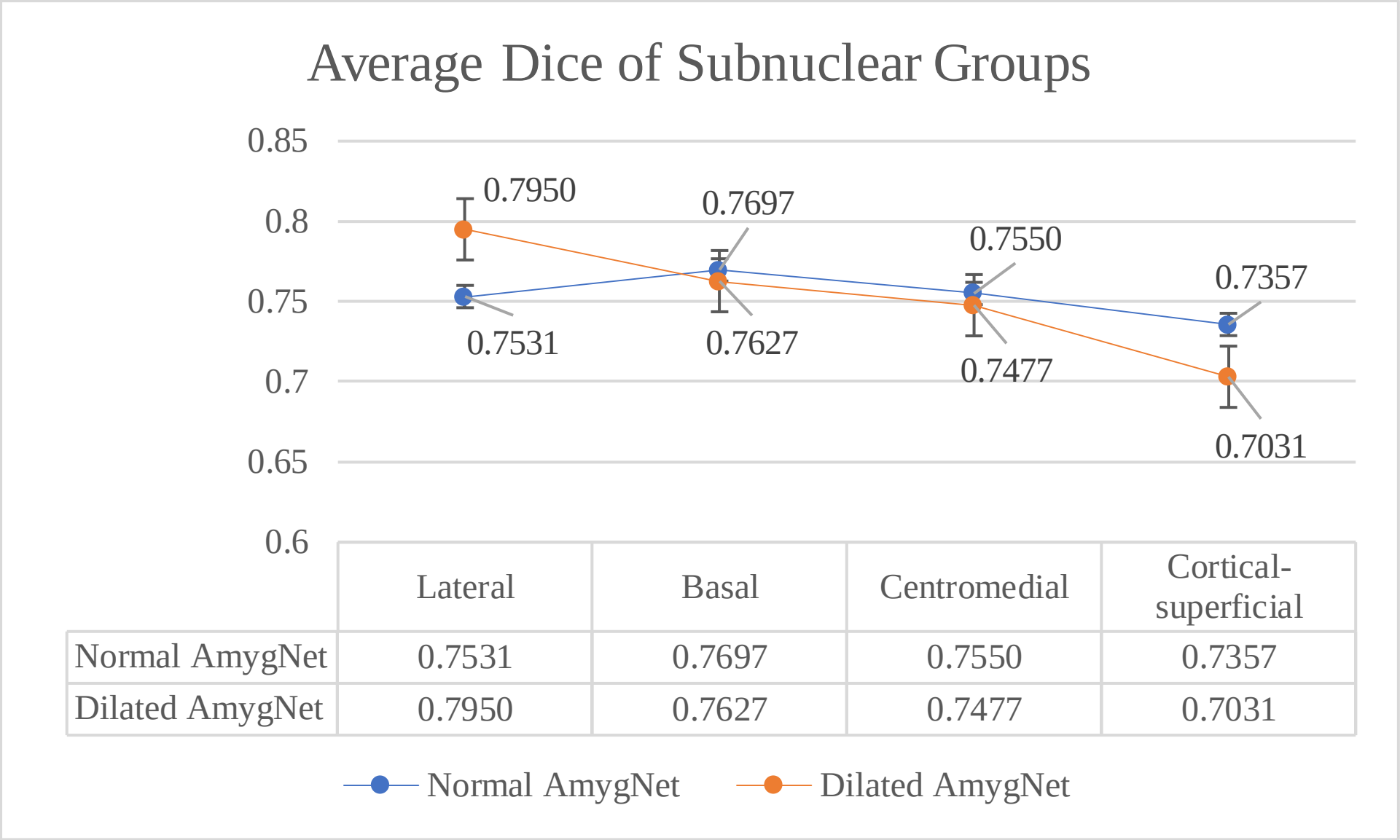

| | Lateral | Basal | Centromedial | Cortical-superficial |
|---|---|---|---|---|
| Normal AmygNet | 0.7531 | 0.7697 | 0.7550 | 0.7357 |
| Dilated AmygNet | 0.7950 | 0.7627 | 0.7477 | 0.7031 |

Figure 3. Evaluate Normal AmygNet and Dilated AmygNet in Dice, based on different subnuclear groups in amygdala. Sorting by volume-to-surface ratio, the largest subnuclear part should be lateral and the second largest should be basal, while centromedial is smaller than basal and cortical-superficial is the smallest subnuclear part. Generally, Normal AmygNet performs better in small objects, while Dilated AmygNet performs better in large objects. From this result, we may conclude that small receptive field is more suitable for segmenting relatively small objects, while large receptive field is more suitable for segmenting relatively large objects.

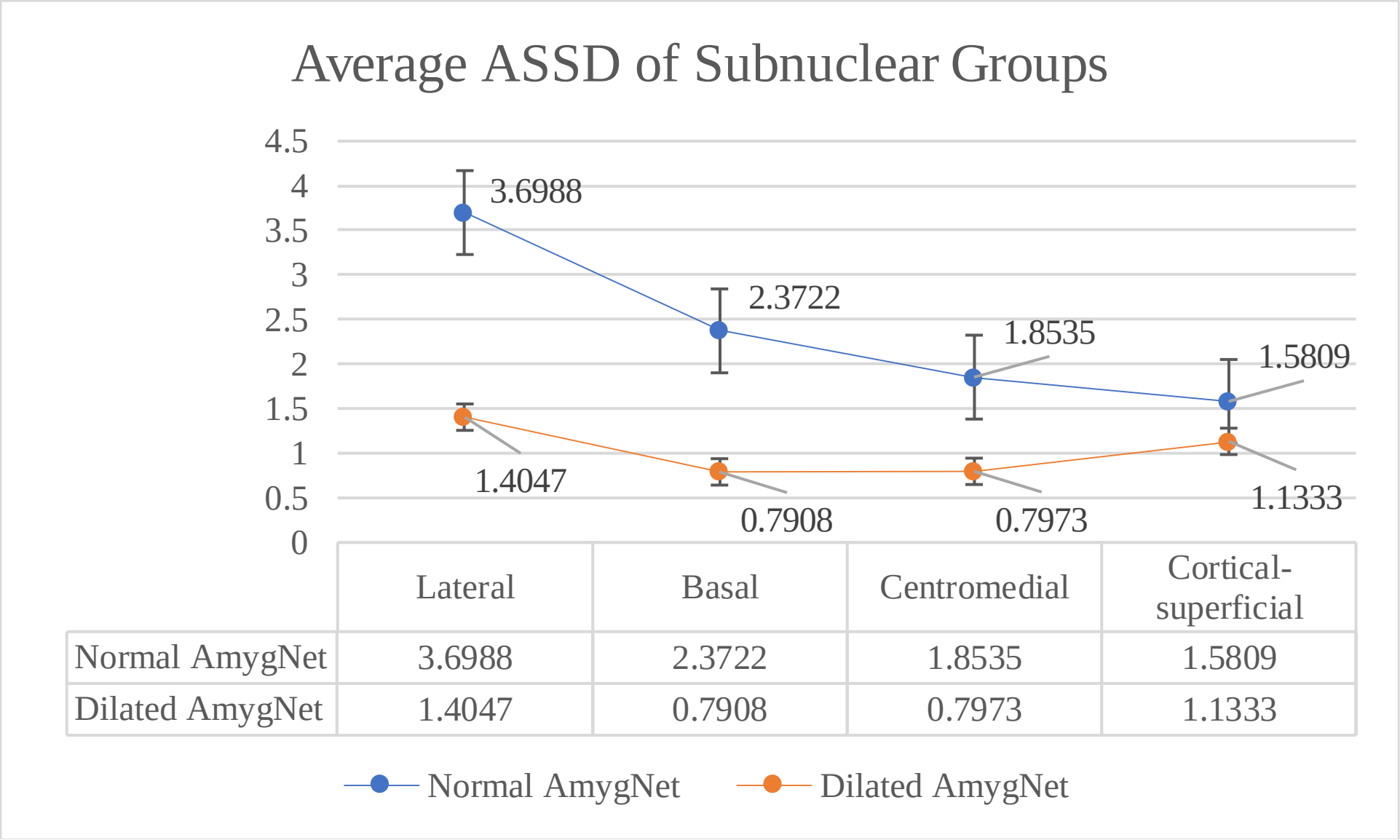

| | Lateral | Basal | Centromedial | Cortical-superficial |
|---|---|---|---|---|
| Normal AmygNet | 3.6988 | 2.3722 | 1.8535 | 1.5809 |
| Dilated AmygNet | 1.4047 | 0.7908 | 0.7973 | 1.1333 |

Figure 4. Evaluate Normal AmygNet and Dilated AmygNet in ASSD, based on different subnuclear groups in amygdala. Sorting by volume-to-surface ratio, the largest subnuclear part should be lateral and the second largest should be basal, while centromedial is smaller than basal

and cortical-superficial is the smallest subnuclear part. Generally, Dilated AmygNet shows predominant advantage, whether the scale of object is large or small. From this result, we may conclude that Dilated AmygNet may perform much better than Normal AmygNet in localization, in both large and small objects.

Classifying results based on subnuclear groups may help us investigate the relation between object scale and receptive field size. Lateral and basal are classified as large-scale objects, while lateral is larger. Compared with lateral and basal, centromedial and cortical-superficial are treated as small-scale objects, while cortical-superficial is smaller. Using Dice to evaluate, normal AmygNet performs better in segmenting small objects, especially in the smallest-object groups (cortical-superficial). Its Dice value exceeds 0.0326 compared with Dilated AmygNet, which is a great gap. However, in segmenting large-object groups, Normal AmygNet performs not that well. In lateral segmentation, which is the largest-object segmentation, Dilated AmygNet achieves 0.0419 higher than Normal AmygNet – this gap is even larger than the gap between Normal AmygNet and Dilated AmygNet in segmenting the smallest subnuclear group, cortical-superficial. While segmenting relatively medium-size objects – basal and centromedial – these two networks provide very close segmenting results – although Normal AmygNet is a little bit better, gaps are both less than 0.01 (0.007 and 0.0073). This trivial difference shows their ability to segment medium-size objects is close. According to Figure 4, Dilated AmygNet achieves more excellent results with no doubt. Dilated AmygNet performs better than Normal AmygNet in each subnuclear group and all 4 ASSD results of Dilated AmygNet are around 1, which is quite competitive compared with Normal AmygNet's results, whose ASSD results are all higher than 1.5.

From the discovery above, we may infer that small receptive field is more suitable for segmenting small objects, while large receptive field is more suitable for segmenting large objects. Normal AmygNet, or smaller receptive field, is extremely incapable in localizing objects, nevertheless, network with larger receptive field is more suitable for localizing. These experiment results motivate us to create a new dual-branch network with two different sizes of receptive fields, which, we suppose, can inherit merits of both small receptive field and large receptive field, and evade drawbacks of these two receptive fields, and finally achieves more outstanding segmentation results in both accuracy and localization.

## 3. AMYGNET

In this section, we will discuss the structure of AmygNet. We introduced Normal AmygNet and Dilated AmygNet while doing pilot experiment in previous section. Except the size of receptive field, other elements of network are the same. To combine the advantage of small receptive field and large receptive field, we propose a new dual-branch network – AmygNet, which has small receptive fields and dilated receptive fields in two separate branches respectively. Consistent with Normal AmygNet and Dilated AmygNet, in each branch of AmygNet, we introduce four ResNet building blocks, with residual connection in each block. One branch uses same parameter settings as Normal AmygNet does, which means receptive fields in it are small (non-dilated), while the other branch uses same parameter settings as Dilated AmygNet does, implementing large receptive fields (dilated). By designing two channels, an input image is able to be segmented by two networks with different sizes of receptive fields. However, these two channels are not isolated, some special manipulation should be conducted to let two channels' data "communicate" with each other. In this experiment, we use a method called "element-wise summation". Basically, it is a process adding up two channels' results into one parameter every time after both channels finish one convolutional manipulation respectively, so in our AmygNet, totally 8 times of element-wise summation are performed (2 times in each ResNet building block). By conducting this operation, a converged result may be presented, which may include both advantages of small receptive field and large receptive field, and make network be outstanding in segmenting objects in various scales.

Figure 5 shows workflow of AmygNet. Upper branch implements small receptive fields, and lower branch implements large receptive fields. In upper branch, when implementing ResNet, each convolution implements a kernel size of 3×3×3, except reshape convolution, which has a kernel size of 1×1×1 (used for residual connection). Stride in each convolution is same, too, which has a value of 1. Since receptive fields in this branch are non-dilated, dilation rate of each convolution is 1. Before 4 layers of ResNet blocks, there is a convolution layer, which has kernel size of 3×3×3, outputs 30 channels. This sets the number of input channels of first ResNet block is 30. After four ResNet blocks, two full connected layers are implemented before generating the results. Lower branch implements exactly the same parameter settings as upper branch, except dilation rate in each convolution. In conv1 stage, dilation rate is 2 and 4 in first and second convolution respectively. In conv2 stage, dilation rate is 2 and 8 in first and second convolution respectively. In conv3 stage, dilation rate is 2 and 4 in first and second convolution respectively. In conv4 stage, dilation rate is 2 and 1 in first and second convolution respectively.

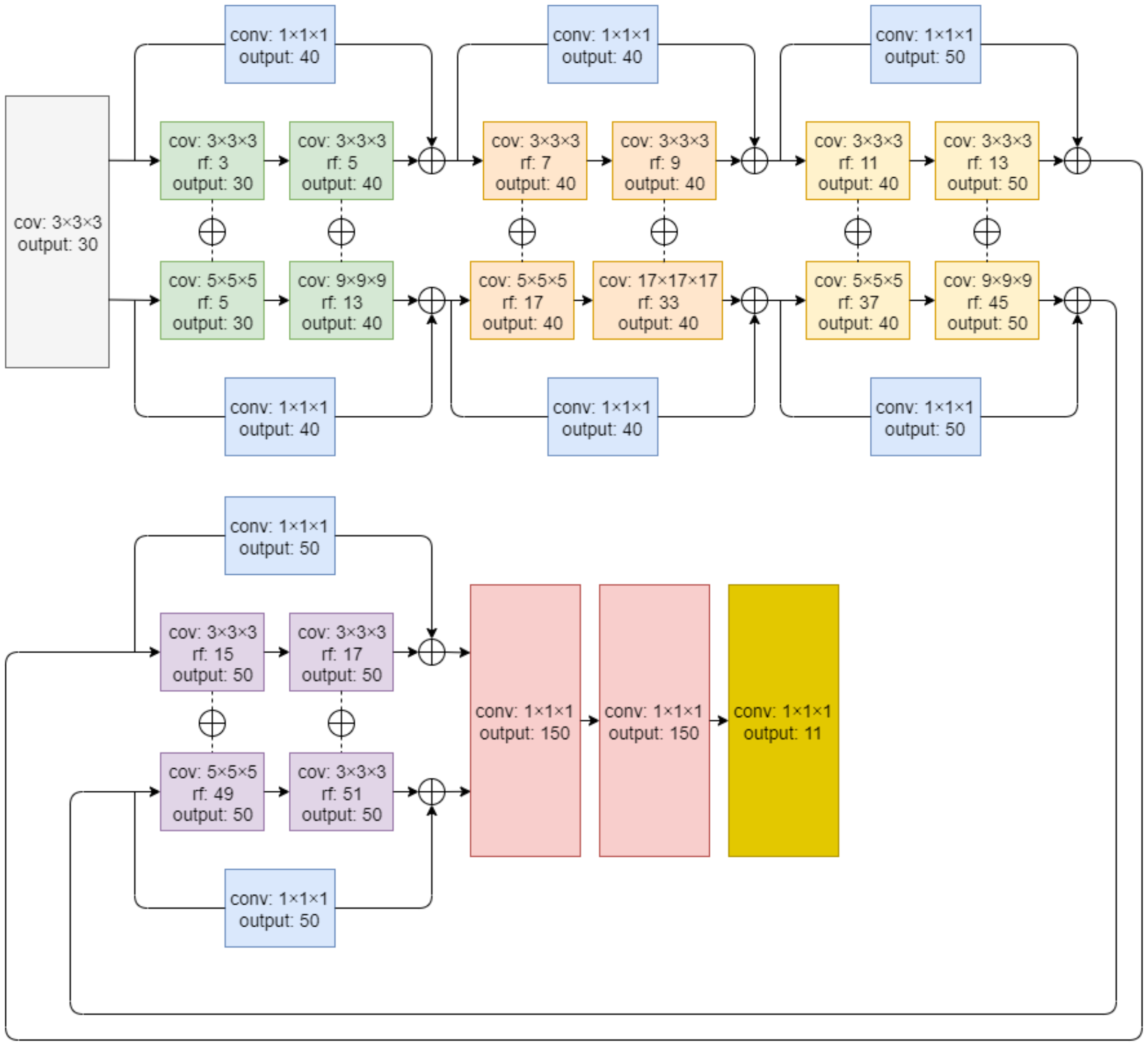

Figure 5. The structure of AmygNet. Four ResNet building blocks are implemented in upper and lower branches. “conv” represents size of convolutional kernels, while “rf” means the size of receptive field at that layer and “output” shows the output size of that layer. Upper branch uses small (non-dilated) receptive fields, while lower branch uses large (dilated) receptive fields.

Element-wise summation is used to converge results of two branches' convolutions in each layer, presenting by connecting one convolution with its corresponding convolution in the other branch. First layer and two full connected layers are also shown at the first and last stage of experiment respectively.

## 4. EXPERIMENT

In this section, we will present our experiment process. Main theory of our experiment is given a sample of whole human brain, using AmygNet to segment subnuclear groups in each amygdala. Since there are differences between different subnuclear groups' sizes in one amygdala, we may record results of each subnuclear group to see if AmygNet can perform well in segmenting objects in various scales.

Based on two pilot experiments, we notice small receptive field may be good for segmenting small objects, while large receptive field may be good for segmenting large objects. We suppose that by combining two sizes of receptive fields, dual-branch AmygNet will be an outstanding network that can segment various sizes of objects well. In this experiment, process and parameter setting are the same as pilot experiments. That is, using the same dataset (14 3-dimentional MRI human brain images), focusing on amygdalae and their subnuclear groups (lateral, basal, centromedial and cortical-superficial) and doing the experiment in the same way. AmygNet is trained in 11 patches on Google Colaboratory, using 1 GPU. Number of classes is also set to 11 (although we only concern about 8 classes). In total, we train 500 epochs. AmygNet is trained, validated and tested on 10 training samples, 2 validation samples and 2 testing samples respectively. To make investigation thoroughly, we set each sample as testing sample once while evaluating AmygNet, which means each class will have 14 results after finishing the experiment. If 2 samples will be tested in each running and there are 14 samples in total, theoretically, 7 times of running (including training, validation, testing and evaluation in each running) will be needed. Since we set the number of samples in validation set and the number of samples in testing set both equal to 2, to reduce training time, after one running finished, samples in validation set and in testing set will be swapped – validation set in last round becomes testing set for next round, and testing set in last round becomes validation set for nest round – then another running may start from validation part (training part is not needed, because training sets of these two rounds are the same), followed by testing and evaluation. In other words, a trained output will be used twice, so training time is halved. Table 2 shows how we arrange our dataset to achieve experiment efficiently.

Table 2. Training set, validation set and testing set setups for evaluating an AmygNet. By swapping samples in validation set and samples in testing set after 1st running, 3rd running and 5th running respectively, 2nd running, 4th running and 6th running do not need to train training set again, and can start from validation step. Thus, only 4 times of training is needed, instead of 7 times (1 training for each running) and time of running 3 training sets is saved.

<table>
<tr><th>Running No.</th><th>Training Set</th><th>Validation Set</th><th>Testing Set</th></tr>
<tr><td>1</td><td rowspan="2">subject209, subject211, subject212, subject213, subject214, subject215, subject217, subject218, subject219, subject220</td><td>subject205, subject206</td><td>subject207, subject208</td></tr>
<tr><td>2</td><td>subject207, subject208</td><td>subject205, subject206</td></tr>
</table>

| | | | |
|---|---|---|---|
| 3 | subject205, subject206, subject207, subject208, subject214, subject215, subject217, subject218, subject219, subject220 | subject209, subject211 | subject212, subject213 |
| 4 | | subject212, subject213 | subject209, subject211 |
| 5 | subject205, subject206, subject207, subject208, subject209, subject211, subject212, subject213, subject214, subject215 | subject217, subject218 | subject219, subject220 |
| 6 | | subject219, subject220 | subject217, subject218 |
| 7 | subject205, subject206, subject207, subject208, subject209, subject211, subject212, subject213, subject214, subject215 | subject212, subject213 | subject214, subject215 |

Above experiment method also applies in pilot experiments.

## 5. Results

In this section, some results and evaluations of these results will be presented. Basically, we use Dice and ASSD to evaluate AmygNet as we did in pilot experiments. We generate results based on experiment results of 8 classes (see Table 1). Since every sample is set as testing sample once and we have 14 samples in data set, 14 results are generated per class. Averaging 14 results in each class and combining the results of pilot experiments, we get Figure 6 and Figure 7.

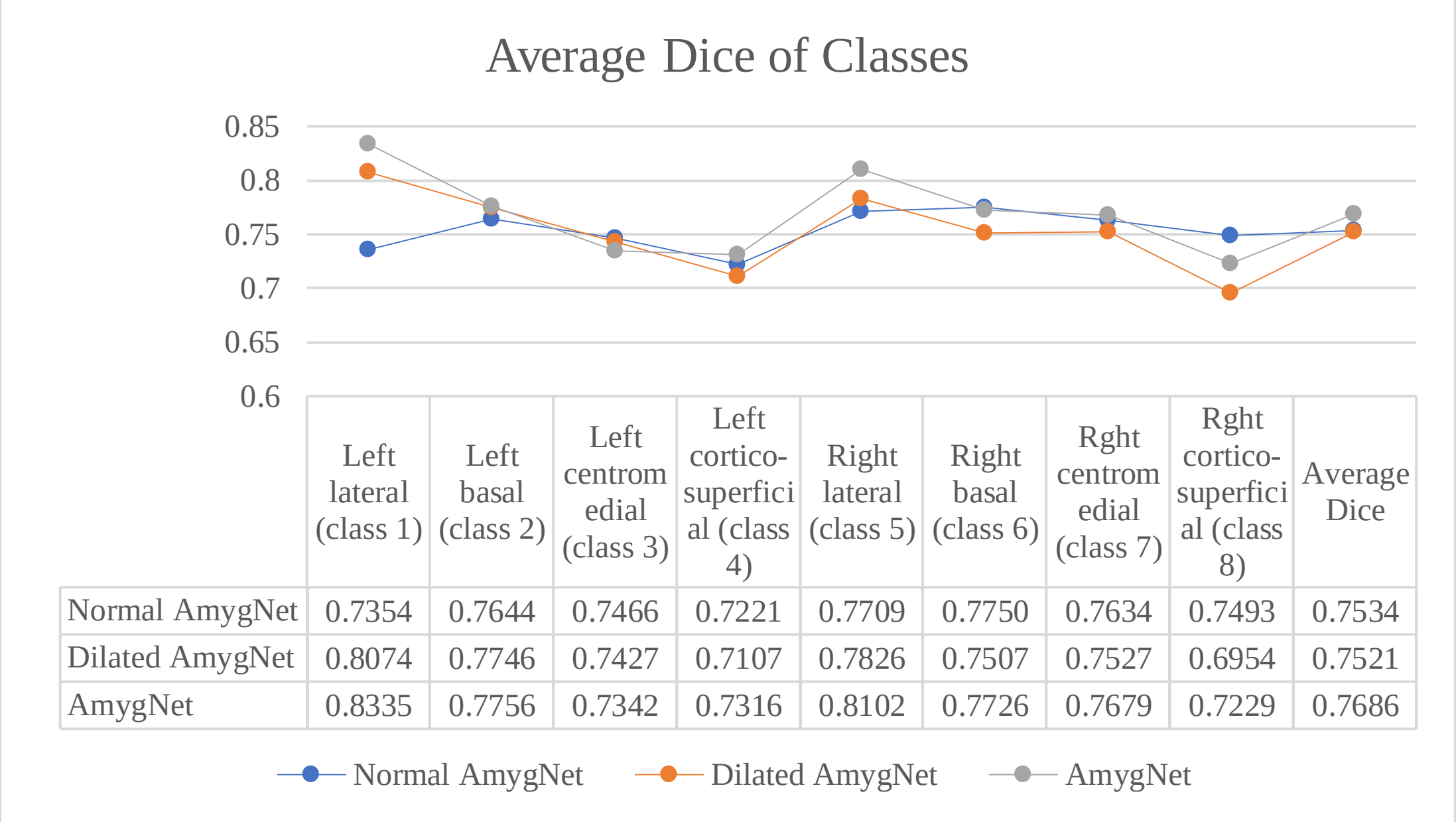

| | Left lateral (class 1) | Left basal (class 2) | Left centromedial (class 3) | Left cortico-superficial (class 4) | Right lateral (class 5) | Right basal (class 6) | Rght centromedial (class 7) | Rght cortico-superficial (class 8) | Average Dice |
|---|---|---|---|---|---|---|---|---|---|
| Normal AmygNet | 0.7354 | 0.7644 | 0.7466 | 0.7221 | 0.7709 | 0.7750 | 0.7634 | 0.7493 | 0.7534 |
| Dilated AmygNet | 0.8074 | 0.7746 | 0.7427 | 0.7107 | 0.7826 | 0.7507 | 0.7527 | 0.6954 | 0.7521 |
| AmygNet | 0.8335 | 0.7756 | 0.7342 | 0.7316 | 0.8102 | 0.7726 | 0.7679 | 0.7229 | 0.7686 |

Figure 6. Average Dice of different classes using Normal AmygNet (from pilot experiments), Dilated AmygNet (from pilot experiments) and AmygNet. Using both small receptive field and large receptive field, AmygNet achieves best Dice in 5 classes among these three networks and obtains best average Dice. This result shows applying different sizes of receptive fields in one

network may achieve better segmentation accuracy than only applying one single size of receptive field.

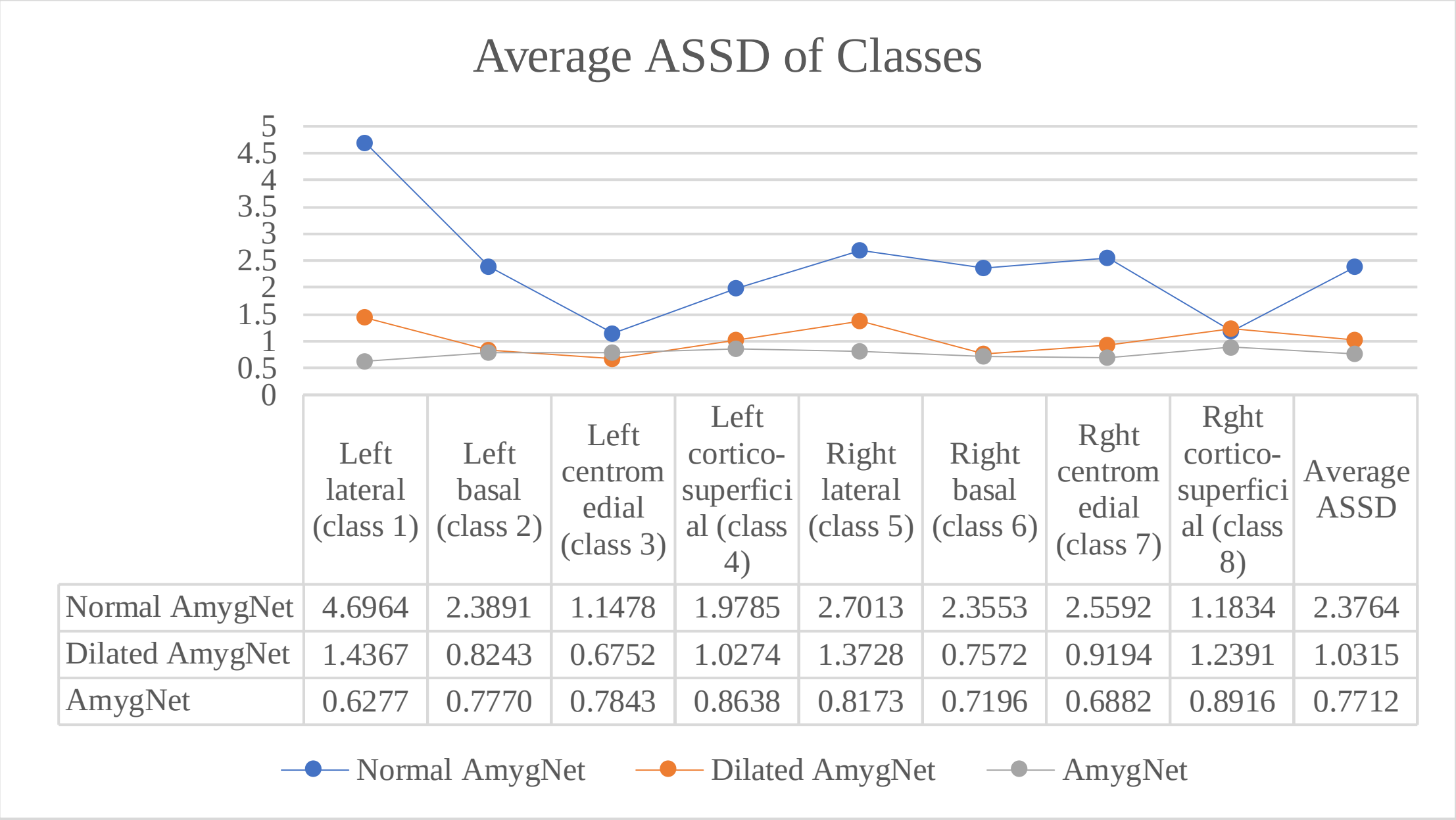

| | Left lateral (class 1) | Left basal (class 2) | Left centromedial (class 3) | Left cortico-superficial (class 4) | Right lateral (class 5) | Right basal (class 6) | Rght centromedial (class 7) | Rght cortico-superficial (class 8) | Average ASSD |
|---|---|---|---|---|---|---|---|---|---|
| Normal AmygNet | 4.6964 | 2.3891 | 1.1478 | 1.9785 | 2.7013 | 2.3553 | 2.5592 | 1.1834 | 2.3764 |
| Dilated AmygNet | 1.4367 | 0.8243 | 0.6752 | 1.0274 | 1.3728 | 0.7572 | 0.9194 | 1.2391 | 1.0315 |
| AmygNet | 0.6277 | 0.7770 | 0.7843 | 0.8638 | 0.8173 | 0.7196 | 0.6882 | 0.8916 | 0.7712 |

Figure 7. Average ASSD of different classes using Normal AmygNet (from pilot experimetns), Dilated AmygNet (from pilot experiments) and AmygNet. Using both small receptive field and large receptive field, AmygNet achieves best ASSD in 7 classes among these three networks and obtains best average ASSD. This result shows applying different sizes of receptive fields in one network may achieve better localization ability in boundary and regions of interests.

From two figures above, it is not hard to see that AmygNet performs better in both Dice and ASSD evaluation criteria. From the perspective of Dice, although Normal AmygNet performs the best in class 3 (centromedial on left side), class 6 (basal on right side) and class 8 (cortico-superficial on right side), AmygNet performs the best in other 5 classes and obtains the best average result overall – 0.0152 better than Normal AmygNet and 0.0165 better than Dilated AmygNet, which are considerable advantages. While evaluating using ASSD, the preponderance of dual-branch network is even greater. Only Dilated AmygNet achieves better than AmygNet in class 3 (centromedial on left side), in other 7 classes and overall average, AmygNet is the most outstanding one without doubt. In overall ASSD, AmygNet is 0.2603 better than Dilated AmygNet and 1.6052 better than Normal AmygNet, which shows AmygNet's unparalleled potential in localizing boundary and regions of interests.

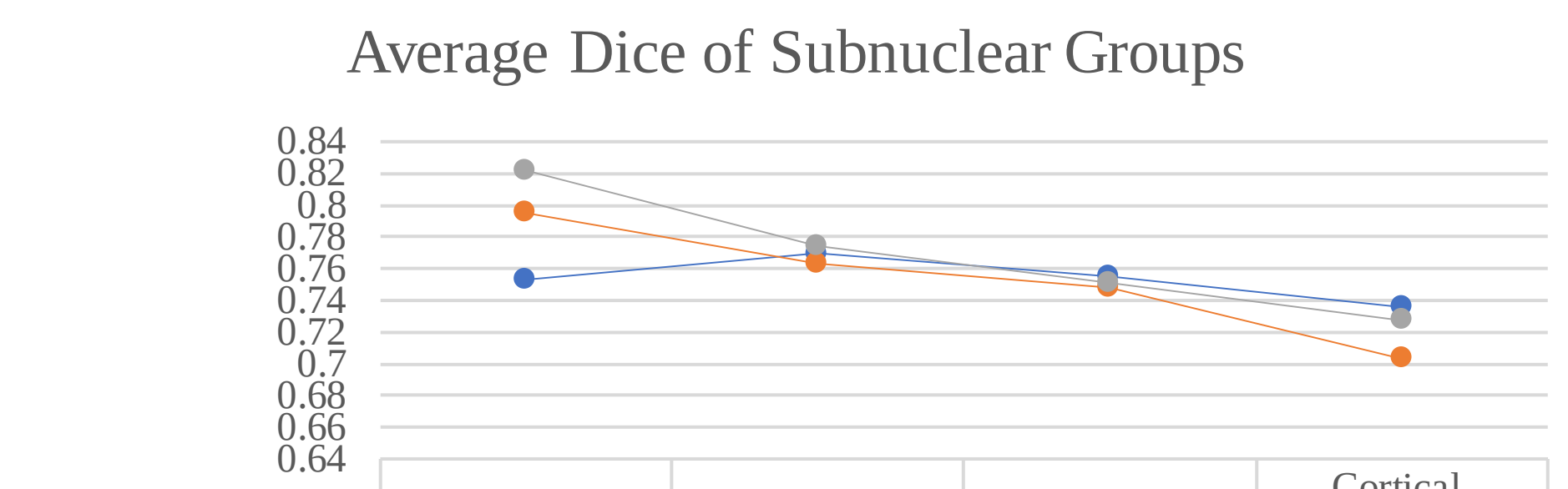

| | Lateral | Basal | Centromedial | Cortical-superficial |
|---|---|---|---|---|
| Normal AmygNet | 0.7531 | 0.7697 | 0.755 | 0.7357 |
| Dilated AmygNet | 0.795 | 0.7627 | 0.7477 | 0.7031 |
| AmygNet | 0.8218 | 0.7741 | 0.751 | 0.7273 |

Normal AmygNet — Dilated AmygNet — AmygNet

Figure 8. Evaluate AmygNets in Dice, based on different subnuclear groups in amygdalae. AmygNet performs the best in segmenting relatively big objects (lateral and basal). Although AmygNet is not the best one in segmenting relatively small objects (centromedial and cortical-superficial), gaps between the results of AmygNet and the best results are both less than 0.01 (0.0084 and 0.004). Overall, AmygNet, using receptive fields in different sizes, performs better than networks using receptive fields in single size.

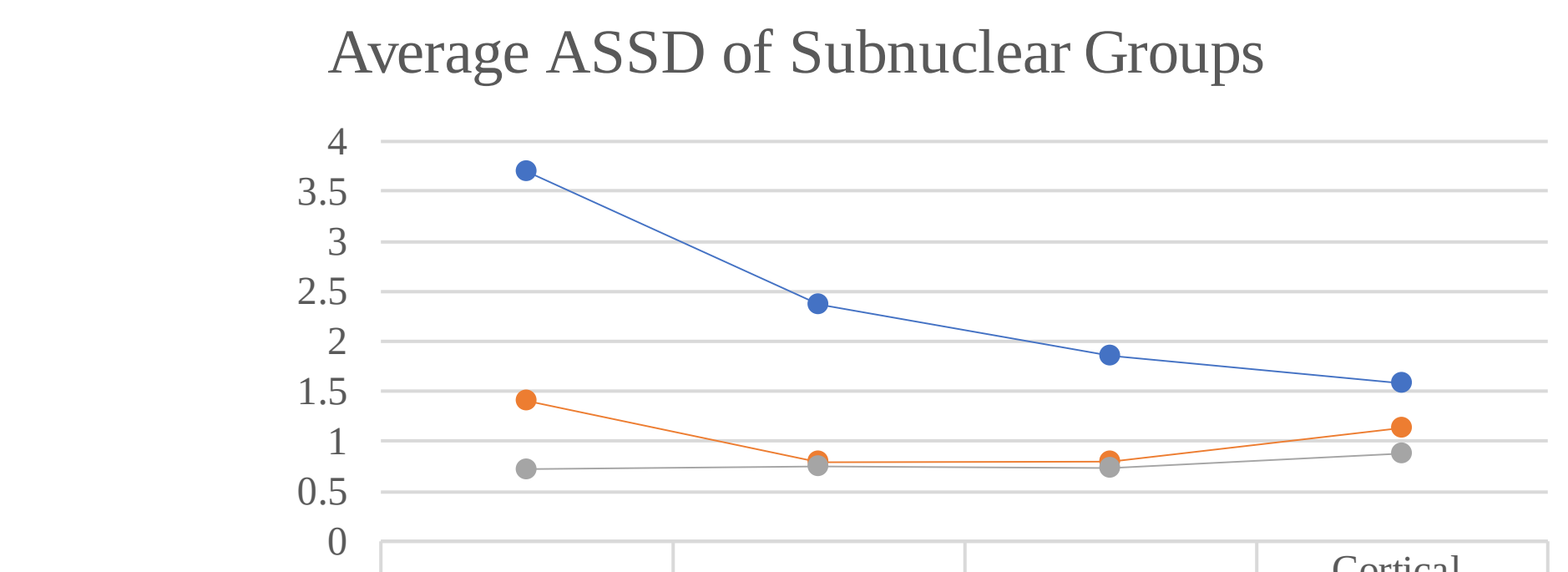

| | Lateral | Basal | Centromedial | Cortical-superficial |
|---|---|---|---|---|
| Normal AmygNet | 3.6988 | 2.3722 | 1.8535 | 1.5809 |
| Dilated AmygNet | 1.4047 | 0.7908 | 0.7973 | 1.1333 |
| AmygNet | 0.7225 | 0.7483 | 0.7363 | 0.8777 |

Normal AmygNet — Dilated AmygNet — AmygNet

Figure 9. Evaluate AmygNets in ASSD, based on different subnuclear groups in amygdalae. ASSDs of Normal AmygNet are all higher than 1.5, which demonstrates small receptive field's incapability in localization. Dilated AmygNet achieves better ASSDs than Normal AmygNet, which shows larger receptive field may be more suitable for localization. However, ASSDs of Dilated AmygNet may fluctuate based on different scales of object. Nevertheless, AmygNet solves this issue. Keeping ASSD always lower than 1, AmygNet, using receptive fields in different sizes, performs best in localization.

By averaging the results of two classes of the same subnuclear group, we may find the results based on subnuclear groups. Figure 8 and Figure 9 show results from the aspect of different subnuclear groups. In relatively large objects (lateral and basal), AmygNet is the best choice (0.0268 and 0.0047 better than the second bests). In segmenting relatively small objects (centromedial and corical-superficial), AmygNet is still competitive – almost as good as Normal AmygNet (only 0.0084 and 0.004 lower than Normal AmygNet). Nevertheless, while using ASSD to evaluate, Normal AmygNet shows its limitation – all ASSDs are higher than 1.5. This result reflects using small receptive field to localize regions of interests may generate result with many isolated regions, which means network with small receptive field may not be capable of performing localization accurately. This trait of small receptive field can also be proved by visualization result (Figure 10). Dilated AmygNet performs better than Normal AmygNet, which manifests large receptive field may be more capable in localization than small receptive field. However, ASSD of lateral and ASSD of cortical-superficial groups are higher than 1 using Dilated AmygNet, or large receptive field. This may indicate the instability of Dilated AmygNet in localization. However, AmygNet shows its undoubtable advantage in localization. Using dual-branch structure and different scales of receptive fields, AmygNet keeps all ASSDs lower than 0.9, which is the level networks using single size of receptive field cannot reach. This outstanding result also shows implementing both small and large receptive fields may be able to localize object more accurately, regardless of its scale.

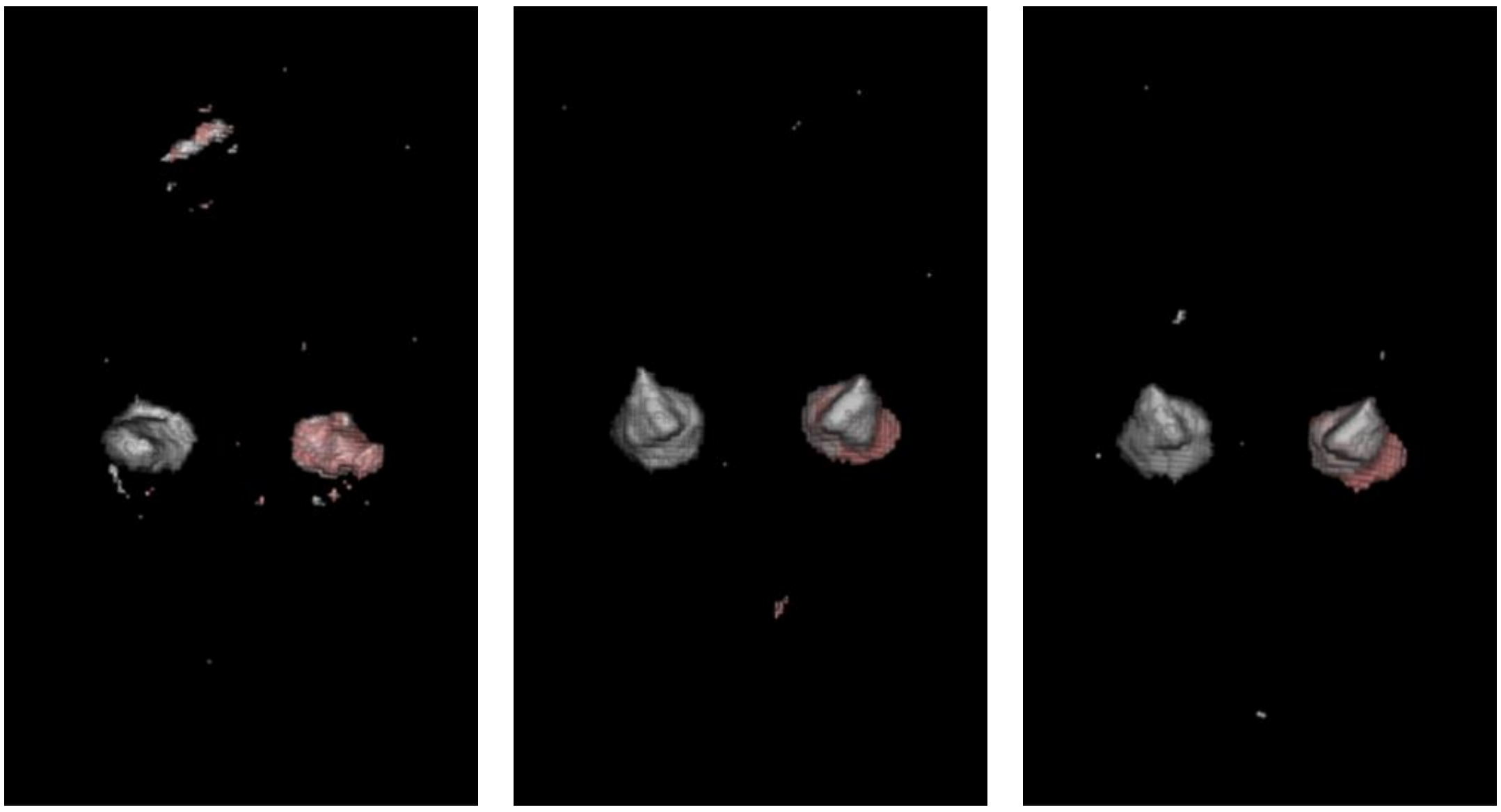

(a) Using Normal AmygNet (b) Using Dilated AmygNet (c) Using of AmygNet

Figure 10. Visualization of segmentation results of one sample. (a) is generated by Normal AmygNet, (b) is generated by Dilated AmygNet and (c) is generated by AmygNet. We may find that (a) has much more isolated regions than (b) and (c), which proves the incapability of small receptive field in localization. This feature can also be reflected in small receptive field having higher ASSD.

Figure 10 shows the segmentation results using different AmygNets (different receptive field combination). There is no doubt Figure 10(a) has more isolated regions than Figure 10(b) and Figure 10(c). These false positive regions may result from small receptive field's worse localization ability, which means the incapability of small receptive field in localization may result in false positive regions in final segmentation result. This result proves small receptive field may not be suitable for localization, or localizing regions of interests, which can be reflected in small receptive field having higher ASSD as well. On the contrary, networks with larger receptive fields (AmygNet and Dilated AmygNet) perform better in localization and have fewer isolated regions. Their lower ASSDs also demonstrate this trait.

## 6. CONCLUSION

In this paper, a dual-branch object segmentation network – AmygNet – is presented to achieve higher segmentation accuracy and better localization ability in segmenting amygdalae, using different sizes of receptive fields. By doing pilot experiments, we find small receptive field may segment small objects better and not be suitable for localization, while large receptive field may segment large objects better and perform well in localization. Based on this, we create AmygNet, using both small receptive field and large receptive field in segmenting objects. Our experiment shows AmygNet takes advantages of both small receptive field and large receptive field, and can achieves better accuracy and localization results in various sizes of objects, which matches our assumption.

On the basis of AmygNet, we considered some possible further work. One possible investigation point is to explore the relation of sizes of receptive fields in two branches. Currently we only conducted experiment using non-dilated receptive fields for small-receptive-fields branch, and using dilated receptive fields with specific dilation rate in each layer for large-receptive-fields branch. Implementing different combinations of dilation rates in both branches may result in totally different experiment results. Another angle is about the kernel size. In our experiment, we use kernel size of $3\times3\times3$ as standard kernel size. For subjects in different scales, standard kernel size may need to be adjusted to adapt change in object size. Experiment related to depth of network may also worth conducting. Deeper network may generate better results, and it may also result in overlooking small objects and inaccurate segmentation, as well as more computation cost. Finding a trade-off point of depth of network is probably a significant factor influencing experiment results.

We believe that AmygNet has great potential in object segmentation, and could lead to a new research direction in medical area. In the future, we would like to explore more in this direction.